\documentclass{emulateapj}
\usepackage{apjfonts,lscape}

\journalinfo{Astrophysical Journal Letters, in press}
\slugcomment{}

\shorttitle{Lyman Alpha Emitters at $z=3.1$}
\shortauthors{Gawiser et al.}

\begin{document}

\title{The Physical Nature of Lyman Alpha Emitting Galaxies at 
$z=3.1$
}

\author{Eric Gawiser\footnotemark[1,2,3,4],
Pieter G. van Dokkum\footnotemark[1,2],
Caryl Gronwall\footnotemark[5], 
Robin Ciardullo\footnotemark[5], 
Guillermo A. Blanc\footnotemark[3],
Francisco J. Castander\footnotemark[6,3],
John Feldmeier\footnotemark[7,4],
Harold Francke\footnotemark[3,1],
Marijn Franx\footnotemark[8],
Lutz Haberzettl\footnotemark[9],
David Herrera\footnotemark[1,2,10],
Thomas Hickey\footnotemark[5], 
Leopoldo Infante\footnotemark[11],
Paulina Lira\footnotemark[3], 
Jos\'e Maza\footnotemark[3],
Ryan Quadri\footnotemark[1],
Alexander Richardson\footnotemark[1,2],
Kevin Schawinski\footnotemark[12],
Mischa Schirmer\footnotemark[13],
Edward N. Taylor\footnotemark[8],
Ezequiel Treister\footnotemark[3,1,2],
C. Megan Urry\footnotemark[2,10], 
Shanil N. Virani\footnotemark[1,2]
\vspace{-0.2 in}
}

\affiliation{}

\footnotetext[1]{Department of Astronomy, Yale University, P.O. Box 208101, New Haven, CT  06520; gawiser,dokkum,herrera,quadri@astro.yale.edu,
alexander.richardson,shanil.virani@yale.edu.}
\footnotetext[2]{Yale Center for Astronomy \& Astrophysics, Yale University, 
P.O. Box 208121, New Haven, CT 06520; meg.urry@yale.edu.}
\footnotetext[3]{Departamento de Astronom\'{\i}a, Universidad de Chile, 
 Casilla 36-D, Santiago, Chile; gblancm,hfrancke,plira,jose,etreiste@das.uchile.cl.}
\footnotetext[4]{National Science Foundation
   Astronomy and Astrophysics Postdoctoral Fellow.}
\footnotetext[5]{Department of Astronomy and Astrophysics, Pennsylvania 
State University, University Park, PA 16802; caryl,rbc,tomhickey@astro.psu.edu.}
\footnotetext[6]{Institut d'Estudis Espacials de Catalunya/CSIC,Gran Capit\`a 2-4, E-08034 Barcelona, Spain; fjc@ieec.fcr.es.}
\footnotetext[7]{National Optical Astronomy Observatories, P.O. Box 26732, 
Tucson, AZ  85726; jjfeldmeier@ysu.edu.}
\footnotetext[8]{Leiden Observatory, Postbus 9513, NL-2300 RA Leiden, Netherlands; franx,ent@strw.leidenuniv.nl.}
\footnotetext[9]{Astronomisches Institut der Ruhr-Universit\"{a}t-Bochum, Universit\"{a}tsstr. 150, 44780 Bochum, Germany; lutz.haberzettl@astro.ruhr-uni-bochum.de.}
\footnotetext[10]{Department of Physics, Yale University, PO Box 208120, New Haven, CT  06520.}
\footnotetext[11]{Departamento de Astronom\'{\i}a y Astrof\'{\i}sica, 
Pontificia Universidad Cat\'olica de Chile, Casilla 306, 22 Santiago, Chile; 
linfante@astro.puc.cl.}
\footnotetext[12]{University of Oxford, Astrophysics, Keble Road, Oxford OX1 3RH; kevins@astro.ox.ac.uk.}
\footnotetext[13]{Institut f\"{u}r Astrophysik und Extraterrestricsche Forschung (IAEF), Universit\"{a}t Bonn, Auf dem H\"{u}gel 71, 53121 Bonn, Germany; 
mischa@ing.iac.es.}

\vspace{-0.2 in}

\begin{abstract} 
We selected 40 candidate  
Lyman Alpha Emitting galaxies (LAEs) at $z\simeq 3.1$ 
with observed frame equivalent widths $>$150\AA\ 
and inferred emission line fluxes
$>2.5\times10^{-17}$ ergs cm$^{-2}$ s$^{-1}$
from deep narrow-band 
and broad-band MUSYC images of the Extended Chandra Deep Field South. 
Covering 992 arcmin$^2$, 
this is the largest ``blank field'' 
surveyed for LAEs at $z\sim3$, allowing an improved 
estimate of the space density of this population of 
$3\pm1\times10^{-4}$$h_{70}^3$Mpc$^{-3}$.  
Spectroscopic follow-up of 23 candidates 
yielded 18 redshifts, all at $z\simeq3.1$.  
Over 80\% of the LAEs are dimmer
in continuum magnitude than the typical Lyman break galaxy 
spectroscopic limit of $R= 25.5$ (AB), with a median 
continuum magnitude $R\simeq 27$ 
and very blue continuum colors,
$(V-z)\simeq 0$.  
Over 80\% of the LAEs have the right $UVR$ colors to 
be selected as Lyman break galaxies, but only 10\% 
also have $R\leq 25.5$.
Stacking the $UBVRIzJK$ fluxes reveals that LAEs have  
stellar masses 
$\simeq5\times10^8 h_{70}^{-2}$ M$_\odot$ and minimal dust extinction, $A_V\la 0.1$.   
Inferred star formation rates are  
$\simeq$6$h_{70}^{-2}$M$_\odot$yr$^{-1}$, 
yielding a cosmic star formation rate 
density of $2\times10^{-3}h_{70}$M$_\odot$yr$^{-1}$Mpc$^{-3}$.  
None of our 
LAE candidates show evidence for rest-frame 
emission line equivalent widths 
EW$_{rest}$$>$240\AA\ which might imply a non-standard IMF.  
One candidate is detected by Chandra, implying an AGN fraction 
of $2\pm2$\% for LAE candidate samples.
In summary, LAEs at $z\sim3$ have rapid star formation, 
low stellar mass, little dust obscuration and 
no evidence for a substantial AGN component.\footnote{Based 
on observations obtained with the Magellan telescopes 
at Las Campanas Observatory and at Cerro Tololo Inter-American 
Observatory, a division of the National Optical Astronomy Observatories, 
which is operated by the Association of Universities for Research in 
Astronomy, Inc. under cooperative agreement with the National Science 
Foundation.
}  
\end{abstract}

\keywords{galaxies:high-redshift}

\section{INTRODUCTION}
\label{sec:intro}

Because Lyman $\alpha$ emission is easily quenched by dust, 
the Lyman Alpha Emitting galaxies (LAEs) 
are often characterized as protogalaxies experiencing their
 first burst of star formation \citep{hum96}.  
However, the differing behavior of Lyman $\alpha$ and 
continuum photons encountering dust and neutral gas 
makes it possible for older galaxies to exhibit 
Lyman $\alpha$ emission when morphology and kinematics favor  
the escape of these photons 
\citep[e.g.][]{haimans98}.  
Hence the LAEs could instead represent an older population with actively 
star-forming regions.

  LAEs offer the chance 
to probe the bulk of the high-redshift galaxy luminosity function 
as the strong emission line allows spectroscopic confirmation of 
objects dimmer than the continuum limit $R\leq 25.5$.  
Previous studies of LAEs 
at $z\sim3$ have concentrated on 
known 
overdensities 
\citep{steideletal00,hayashinoetal04,venemansetal05} or 
searches for Lyman Alpha emission near known Damped Lyman Alpha 
absorption systems 
(\citealp{fynboetal03}, see \citealp{wolfegp05} for a review).  
Blank fields, i.e. those not previously known to contain 
unusual objects or overdensities, have been studied at 
$z=3.1$ and 
$z=3.4$, covering  
468 arcmin$^2$ \citep{ciardulloetal02} and 
70 arcmin$^2$ \citep{cowieh98,huetal98}, respectively.   
Significant work has been done in recent years on large blank fields
 at higher redshifts to 
study the LAE luminosity function at 
$z=3.7$ \citep{fujitaetal03}, 
$z=4.5$ \citep{huetal98}, 
$z=4.9$ \citep{ouchietal03,shimasakuetal03}, 
$z=5.7$ (\citealp{martins04};\citealp{malhotrar04} and references therein)  
and $z=6.5$ \citep{malhotrar04}.   
Spectroscopically confirmed samples are small, including
31 LAEs at $z=3.1$ \citep{venemansetal05}, 
18 at $z=4.5$ \citep[LALA,][]{dawsonetal04} 
27 at $z=5.7$ \citep{huetal04,ouchietal05}, 
and 9 at $z=6.6$ \citep{taniguchietal05}.  
The current investigation expands upon the 
blank-field survey of \citet{ciardulloetal02} by covering twice the 
area to a narrow-band detection limit one magnitude deeper.

The study of Lyman Alpha Emitting galaxies at $z\simeq 3.1$ 
is a major goal 
of the Multiwavelength Survey 
by Yale-Chile (MUSYC, \citealp{gawiseretal06a}, 
\url{http://www.astro.yale.edu/MUSYC}). The Extended Chandra Deep 
Field South (ECDF-S) has been targeted with deep narrow-band imaging 
and multi-object spectroscopy, 
complemented by deep broad-band $UBVRIzJK$ and 
public Chandra+ACIS-I imaging.  These multiwavelength data make it 
possible to study the physical nature of 
LAEs and to distinguish star formation from AGN as the source of 
their emission.  
We assume a $\Lambda$CDM cosmology 
consistent with WMAP results \citep{bennettetal03} with 
$\Omega_m=0.3, \Omega_\Lambda=0.7$ and 
$H_0 = 70 h_{70}$ km s$^{-1}$ Mpc$^{-1}$.  
All magnitudes 
are given in the AB95 system \citep{fukugitaetal96}.  

\section{OBSERVATIONS}
\label{sec:obs}

Our narrow-band imaging 
of the ECDF-S
was obtained using the  
NB5000\AA\ filter (50\AA\ FWHM)
with CTIO4m+MOSAIC-II on several nights from 2002 to 2004 for 
a total of 29 hours of exposure time.  
Our  
$UBVRI$ imaging results from combining public images taken with 
ESO2.2m+WFI by the ESO Deep Public Survey  
and COMBO-17 teams 
\citep{erbenetal05,hildebrandtetal05,arnoutsetal01,wolfetal04}.  Our 
$z'$ imaging was taken with CTIO4m+MOSAIC-II on January 
15, 2005.  
Details of our optical images will be presented in E. Gawiser et al. 
(in prep.).  Our $JK$ images of ECDF-S were obtained with 
CTIO4m+ISPI on several nights during 2003-2004 and will be described 
in E.N. Taylor et al. (in prep.).  
The final images 
cover $31.5'\times31.5'=992$ arcmin$^2$ 
centered on the Chandra Deep Field South 
and were processed through 
the MUSYC photometric pipeline to generate 
APCORR (corrected aperture) fluxes and uncertainties as described in  
\citet{gawiseretal06a}.  Table \ref{tab:depths} gives our source 
detection depths.

\begin{table}[h!]
\begin{center}
\caption{5$\sigma$ Point Source Detection Limits for MUSYC ECDF-S Images in 
AB magnitudes.\label{tab:depths}}
\begin{tabular}{ccccccccc}
\tableline\tableline
NB5000 & U & B & V & R & I & z$'$ & J & K\\
\tableline
25.5 & 26.0 & 26.9 & 26.4 & 26.4 & 24.6 & 23.6 & 22.7 & 22.0\\
\tableline
\end{tabular}
\end{center}
\end{table}

Multi-object spectroscopy of 23 LAE candidates 
was performed with the IMACS instrument on the 
Magellan-Baade telescope on Oct. 26-27, 2003, Oct. 7-8, 2004, and 
Feb. 4-7, 2005.  The 300 line/mm grism was used with 1.2$''$ slitlets 
to cover $4000-9000$\AA\ at a resolution of 7.8\AA\ .  
Details of our spectroscopy will be given in P.~Lira et al. (in prep).  

\section{CANDIDATE SELECTION}
\label{sec:colors}

The greatest challenge in selecting Lyman Alpha Emitting galaxies 
at $z\simeq 3.1$ 
is to minimize contamination from $z\simeq 0.34$ 
galaxies exhibiting emission lines in [O II]3727\AA\ .  
These interlopers can be avoided by requiring a 
high equivalent width ($>$150\AA\ in the observed frame) 
which eliminates all but the 
rarest [O II] emitters \citep{terlevichetal91,sternetal00}.   
Contamination from 
[O III]5007\AA\ is minimal at these wavelengths, as the volume for 
extragalactic emitters is small, and Galactic planetary nebulae 
are very rare at such high Galactic latitude ($b=-54$).

Selecting LAEs requires an estimate of the continuum 
at the wavelength of the narrow-band filter, so we tested 
weighted sums of the $B$ and $V$ flux densities and found that 
 $f_\nu^{BV}=(f_\nu^B+f_\nu^V)/2$ minimizes the scatter in 
predicting the NB5000 flux density of typical objects.  
The ``narrow-band excess'' in magnitudes, $BV - NB5000$, was 
then used to select the candidate LAEs
with $(BV - NB5000)>1.5$,  
corresponding to EW$_{obs}^{Ly\alpha}>150$\AA\ . 
When the 
broad-band fluxes are small, 
significant errors in the 
equivalent width estimate may result, and a small fraction 
of the numerous objects without emission lines, 
i.e. with $(BV-NB5000) \simeq 0$,  
could enter the 
``narrow-band excess'' sample.  To avoid both types of interlopers, 
we calculated a formal uncertainty in the $BV-NB5000$ color in 
magnitudes, $\sigma_{BV-NB}$, 
and required candidate LAEs to have ($BV-NB5000$)$ - \sigma_{BV-NB} >1.5$ 
and ($BV-NB5000$)$ - 3 \sigma_{BV-NB} > 0$.      
The latter criterion is similar to the color excess requirement 
of \citet{bunkeretal95}, but our color uncertainties are object-specific 
and account for variation in image depth across the field 
\citep[see][]{gawiseretal06a}.  
To make spectroscopic confirmation feasible, we also required 
NB5000$<$25.0, implying an 
emission line flux $\geq 2.5\times10^{-17}$ ergs cm$^{-2}$ s$^{-1}$.   
Visual inspection to eliminate false narrow-band detections caused by 
CCD defects or cosmic ray residuals resulted in 40 candidate 
LAEs.  23 of these candidates 
have been observed spectroscopically, yielding 
18 confirmations where 
the Lyman $\alpha$ emission line was clearly detected
 in both the two-dimensional 
and extracted spectra and no other emission lines were visible across 
the full optical spectrum.  
We tested the procedure by observing lower-equivalent width objects 
and found several [O II]3727 emitters; all of these interlopers exhibit 
clear emission lines in H$\beta$, [O III]4959,5007 
and H$\alpha$.  Five of the LAE candidate spectra 
fail to show 
emission lines. 

\begin{figure}[h!]
\vspace{-0.2in}
\includegraphics[angle=0,scale=0.35]{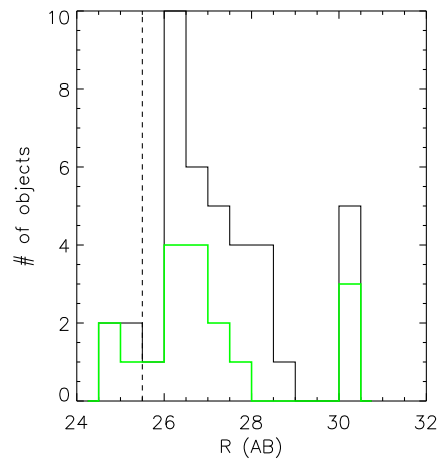}
\caption{
Histogram of $R$-band magnitudes for candidate LAEs (thin 
histogram) and the subset of 
confirmed LAEs (thick histogram), with 
typical 
spectroscopic limit of $R=25.5$ marked with 
dashed vertical line.  Objects with negative $R$ fluxes were 
assigned $R=30$.  
\label{fig:maghist}
}
\end{figure}

\section{RESULTS}
\label{sec:results}

\begin{figure}[h!]
\includegraphics[angle=0,scale=0.38]{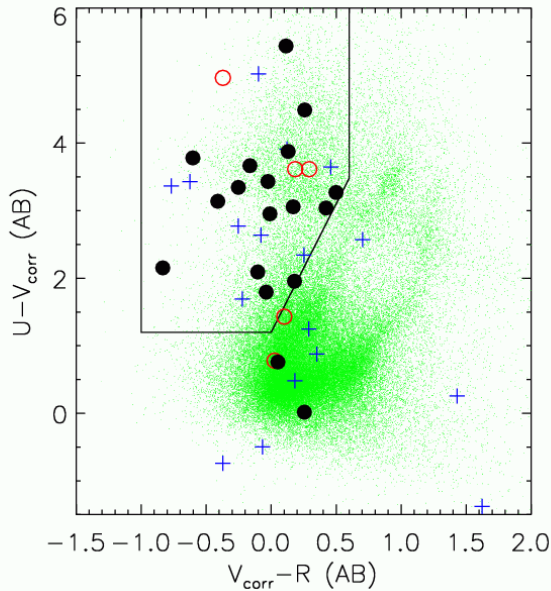}
\caption{
$UVR$ color-color plot of confirmed LAEs (solid circles), 
candidate LAEs with spectroscopy but no confirmed redshift 
(open circles) and candidate LAEs without spectroscopy (plusses)
versus distribution 
of the entire 84,410 object optical catalog (dots).  
The polygonal region in 
the upper left is the Lyman break galaxy
 selection region.
\label{fig:uvr}
}
\end{figure}

Figure \ref{fig:maghist} shows the distribution of candidate 
and confirmed LAE $R$-band continuum magnitudes versus 
 the ``spectroscopic'' Lyman break galaxy (LBG) 
limit of $R\leq 25.5$ \citep{steideletal03}.  Our 
study of LAEs is able to observe objects much dimmer than this, 
with a median 
magnitude $R\sim27$.    
36/40 candidates and 15/18 confirmed LAEs have $R>25.5$, 
showing the efficacy of LAE selection in identifying  
objects from the bulk of the high-redshift galaxy luminosity function.
Figure \ref{fig:uvr} shows 
$UV_{corr}R$ colors of our LAE candidates versus the LBG 
selection region determined by \citet{gawiseretal06a}, 
where $V_{corr}$ refers 
to the $V$-band magnitude after subtracting the flux contributed by the
Lyman $\alpha$ emission lines.
Only 2 out of 18 confirmed 
LAEs fall within the $R\leq25.5$ ``spectroscopic'' 
LBG sample,
%
but 16 out of 18 
fall within 
the color selection region.  About half of our candidate LAEs would meet 
the $R<27$ magnitude limit of the ``photometric'' LBG sample explored 
by \citet{sawickit05}, and these objects should comprise 5\% 
of their sample.    

\vspace{-0.1 in}

\begin{figure}[h!]
\includegraphics[angle=0,scale=0.5]{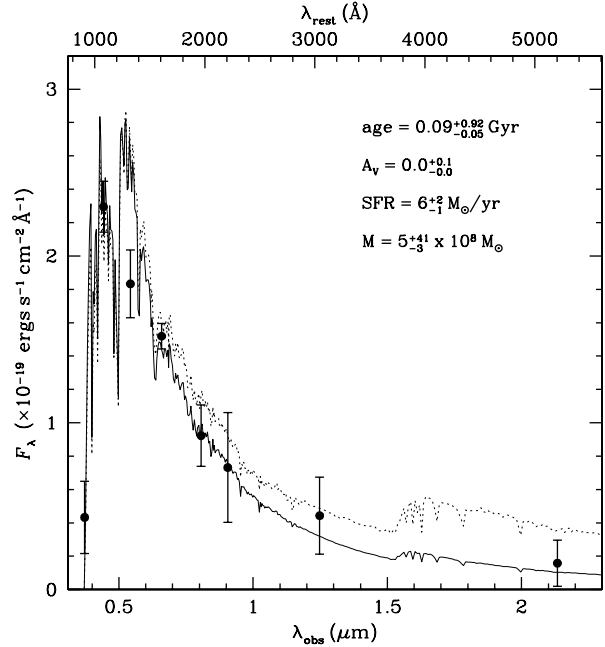}
\caption{
$UBVRIzJK$ broad-band photometry (average flux density of 
 stacked sample) 
of confirmed LAEs 
along with best-fit model from SED fitting (solid) with 
model parameters listed.  The dotted curve shows a maximally 
old model 
with stellar population age fixed to  
2 Gyr (the age of the universe at $z=3.1$), 
$A_V=0.1$, SFR=7$h_{70}^{-2}$ M$_\odot$ yr$^{-1}$ 
and $M_* = 1.1\times 10^{10} h_{70}^{-2}$ M$_\odot$.    
\label{fig:sed}
}
\end{figure}

In order to investigate the full SED of the LAEs, which are too dim 
to obtain individual detections in our NIR photometry, we 
measured stacked fluxes for the confirmed sample 
and show the results of SED modelling in Fig. \ref{fig:sed}.   
\citet{bruzualc03}
population synthesis models were used, with 
constant star formation rate, 
a \citet{salpeter55}
initial mass function from 
0.1M$_\odot$ to 100M$_\odot$, 
solar metallicity  
and \citet{calzettietal97} dust reddening
\citep[e.g.][]{forsterschreiberetal04,vandokkumetal04}.  
Uncertainties in the stacked photometry were determined using bootstrap 
resampling and are close to the formal errors calculated 
from the reported APCORR flux uncertainties.  Parameter uncertainties 
were computed via a Monte Carlo analysis where the stacked fluxes were 
varied within their uncertainties to yield a probability 
distribution of best-fit 
parameters.   
The age of the stellar population is weakly constrained and has been 
restricted to the physically reasonable range 10 Myr $\leq t_* \leq$ 2 Gyr. 
The best-fit parameters shown in  Fig. \ref{fig:sed} 
correspond to minimal dust extinction,   
significant 
star formation rates  
(5$\leq$SFR$\leq$23 $h_{70}^{-2}$ M$_\odot$yr$^{-1}$ 
at 95\% confidence)  
and low stellar mass 
(the 95\% confidence upper limit is $M_*=8.5\times10^9 h_{70}^{-2}$M$_\odot$).
  The LAEs 
appear to have much less dust and stellar mass than 
the $\sim500$ Myr old, $A_V\simeq1$, $\sim2\times10^{10}$M$_\odot$ Lyman break galaxy 
population \citep{shapleyetal01} or the 
$\sim2$ Gyr old, $A_V\simeq2.5$, $\sim10^{11}$M$_\odot$ Distant Red Galaxy 
population \citep{forsterschreiberetal04}.  
The star formation rates of the confirmed LAEs 
inferred 
from their Lyman $\alpha$ luminosities 
average 5$h_{70}^{-2}$ M$_\odot$yr$^{-1}$
and from their rest-frame UV continuum luminosity densities 
average 9$h_{70}^{-2}$ M$_\odot$yr$^{-1}$.    
The consistency of these values with the best-fit SFR from SED modelling 
implies  
minimal dust extinction.

\vspace{-0.1 in}

To check for AGN contamination of our LAE candidate sample, we 
have looked for Chandra detections of these objects.  One LAE candidate 
has an X-ray detection in the catalogs of \citet{viranietal05}
and \citet{lehmeretal05b}, with a 0.5-8keV luminosity of 
10$^{44}$ erg s$^{-1}$. 
No other candidates showed individual detections, so we removed 
this object and performed a stacking 
analysis \citep[e.g.][]{rubinetal04,lehmeretal05a} 
 which resulted in a non-detection of the entire population.  
Using the conversion between SFR and X-ray flux given by \citet{ranallietal03},
the upper limit 
on the average star formation rate per object is 
200$h_{70}^{-2}$ M$_\odot$ yr$^{-1}$, which 
is clearly consistent with the observed SFR.
None of our LAE spectra show broad emission line widths 
($> 1000$ km s$^{-1}$) 
that would be 
inconsistent with the energetics of star formation.  
We therefore expect that very few LAE candidates contain 
luminous AGN which dominate their Lyman $\alpha$ or continuum emission.  

\section{DISCUSSION}
\label{sec:discussion}
Our survey covers $31.5'\times31.5'\times(\Delta z=0.04)$ or 
$59\times59\times38h_{70}^{-3}$Mpc$^3$, yielding an LAE 
number density of 
$3\pm1\times10^{-4}$$h_{70}^3$Mpc$^{-3}$, 
equivalent to $4000\pm 1600$ deg$^{-2}$ per unit 
redshift.
The survey volume was computed using the filter bandpass 
FWHM=50{\AA}, and the five candidates without confirmed redshifts 
were assumed to be LAEs.  
The error bars account for variations in the LAE abundance 
within our survey volume 
caused by large-scale structure assuming a bias of 2.  
The true uncertainties could 
be bigger given the large fluctuations in density observed 
for LAEs at $z=4.9$ by \citet{shimasakuetal04}.
Combining the measured number density and using the best-fit 
star formation rate per object of 6$h_{70}^{-2}$ M$_\odot$ yr$^{-1}$, 
we find a cosmic star formation 
rate density of $2\times10^{-3}h_{70}$ M$_\odot$ yr$^{-1}$Mpc$^{-3}$.  
This is significantly less than the LBG SFR density \citep{steideletal99}, 
but it underestimates the total LAE contribution due to our requirements  
of high 
equivalent width and relatively bright 
$NB5000$ flux designed to select 
a pure sample amenable to spectroscopic confirmation.  
A detailed calculation of the LAE luminosity function at $z\simeq 3.1$, 
which can be integrated to give a fuller estimate of the SFR density, 
will be given in C.~Gronwall et al. (in prep).

The number density, stellar masses, star formation rates, and median 
UV continuum fluxes found for LAEs are within a factor of three of those
predicted by \citet{ledelliouetal05,ledelliouetal06}; the agreement is 
even better when our equivalent width threshold is accounted for.  
The only strong disagreement seen versus these models is their claimed escape 
fraction of 0.02 for Lyman $\alpha$ photons versus our lower limit of 
0.2 (and best-fit of 0.8)  
implied by the comparison of star formation rates determined 
from the observed Lyman $\alpha$ luminosities and SED modelling.  This 
discrepancy could be resolved by using a larger escape fraction and 
a standard IMF instead of the top-heavy IMF assumed in the models.

Our determination that $z=3.1$ LAEs are 
predominantly blue contrasts with the results of 
\citet{stiavellietal01} and \citet{pascarelleetal98}
that LAEs in blank fields at $z\simeq 2.4$ 
are typically red, 
($B-I$)$\simeq1.8$.   
This differs from the median value of ($V_{corr}-z$) $\simeq 0.1$ for 
our spectroscopically confirmed LAEs and the 
median color ($V-I$) $\simeq 0.1$ 
measured by \citet{venemansetal05}.
%
%
%
%
%
%
The difference seems unlikely to be caused by 
evolution in the 
LAE population from $z=3.1$ to $z=2.4$ given the small increase 
in the age of the universe. 

At $z=4.5$, LALA \citep{malhotrar02} reported that a majority of LAE 
candidates had EW$_{rest} > 240$\AA\ , providing   
evidence 
of a top-heavy IMF possibly caused by Population III stars, 
although equivalent widths this high could also result from 
highly anisotropic radiative transfer due to the differing effects 
of dust and gas on Lyman $\alpha$ and UV-continuum photons. 
This photometric measurement 
is sensitive to considerable scatter when the sample is 
selected in the narrow-band and the broad-band imaging 
is shallow, as broad-band non-detections can receive extremely 
large implied equivalent 
widths, and this is guaranteed to occur for any spurious narrow-band 
detections. Indeed, when 
$2\sigma$ upper limits on the continuum flux were used, 
only 10\% of their $z=4.5$ sample had such high EWs, and 
$\sim 20$\% of the confirmed objects have EW$_{rest} > 240$\AA\ 
\citep{dawsonetal04}.  We do not find equivalent widths 
this high for any of our LAE candidates at $z=3.1$.
The difference might reveal evolution in the 
LAE population or could be the result of small number statistics.  

\vspace{-0.2 in}
\acknowledgments

We acknowledge the referee, Andrew Bunker,
 for helpful comments that improved this Letter. 
We thank James Rhoads and Masami Ouchi for valuable conversations.
%
%
We are grateful for 
support from Fundaci\'{o}n Andes, the FONDAP Centro de Astrof\'{\i}sica, 
and the Yale Astronomy Department.
This material is based upon work supported by the National Science 
Foundation under Grant. Nos. AST-0201667,0137927,0071238, and 0302030 
awarded to E.G., C.G., R.C. and J.F. respectively.   
This work was supported in part by NASA grant HST-GO-09525.13-A. 
%
%
%
Facilities:CTIO(MOSAIC II),LCO(IMACS)

\end{document}